\documentclass[apl,twocolumn,preprintnumbers,superscriptaddress,amsmath,amssymb]{revtex4-1}

\usepackage{graphicx}

\usepackage{dcolumn}
\usepackage{color}

\begin{document}

\title{Exciton dynamics and non-linearities in 2D-hybrid organic perovskites}

\author{K. Abdel-Baki}
\author{F. Boitier}
\author{H. Diab}
\author{G. Lanty}
\author{K. Jemli}
\author{F. L\'ed\'ee}
\affiliation{Laboratoire Aim\'e Cotton, CNRS, Univ. Paris-Sud, ENS Cachan, Universit\'e Paris-Saclay, 91405 Orsay Cedex, France}%
\author{D. Garrot}
\affiliation{GEMAC, CNRS, UVSQ, Universit\'e Paris-Saclay, 45 avenue des \'Etats Unis 78035 Versailles cedex, France.}%
\author{E. Deleporte}
\author{J. S. Lauret}
\email{jean-sebastien.lauret@lac.u-psud.fr}
\affiliation{Laboratoire Aim\'e Cotton, CNRS, Univ. Paris-Sud, ENS Cachan, Universit\'e Paris-Saclay, 91405 Orsay Cedex, France}%

\renewcommand{\thefootnote}{\alph{footnote}}

\begin{abstract}
Due to their high potentiality for photovoltaic applications or coherent light sources, a renewed interest in hybrid organic perovskites has emerged for few years. When they are arranged in two dimensions, these materials can be considered as hybrid quantum wells. One consequence of the unique structure of 2D hybrid organic perovskites is a huge exciton binding energy that can be tailored through chemical engineering. We present experimental investigations of the exciton non-linearities by means of femtosecond pump-probe spectroscopy. The exciton dynamics is fitted with a bi-exponential decay with a free exciton life-time of $\sim$100~ps. Moreover, an ultrafast intraband relaxation ($<150$~fs) is also reported. Finally, the transient modification of the excitonic line is analyzed through the moment analysis and described in terms of reduction of the oscillator strength and linewidth broadening. We show that excitonic non-linearities in 2D hybrid organic perovskites share some behaviours of inorganic semiconductors despite their high exciton binding energy.
\end{abstract}

\maketitle
\section{INTRODUCTION}
Hybrid organic-inorganic materials have attracted much attention for few years because of their high potentialities for optoelectronic and photovoltaic applications. The idea is to combine the advantages of organic materials such as the tunability of most of their physical properties,  with the one of inorganic materials such as band engineering or electrical injection. The understanding of the electronic properties of these materials is a key issue to optimize their potential in view of applications. From a fundamental point of view, studying hybrid materials leads to new physics where features of organic and inorganic materials are observed at the same time \cite{Wenus2006,Lanty2011, ForrestPRL2014}.
\begin{figure}[!ht]
\begin{center}
\includegraphics*[scale=.3]{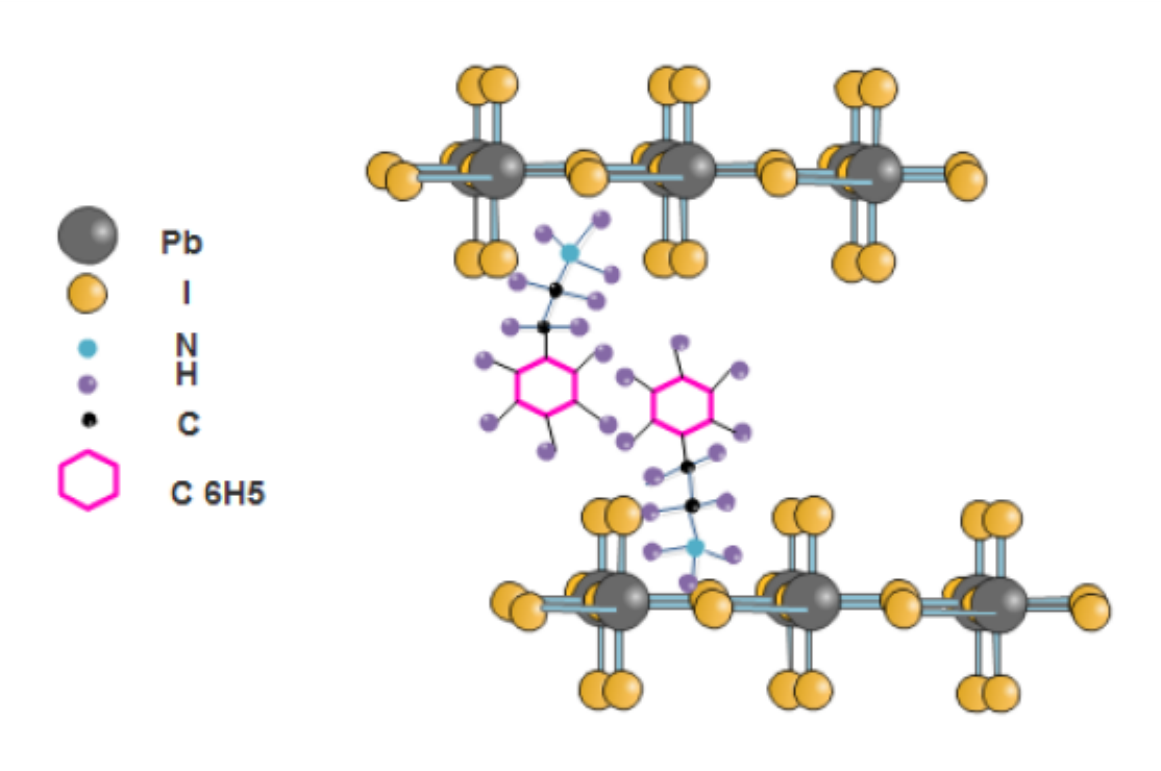}
\includegraphics*[scale=.3]{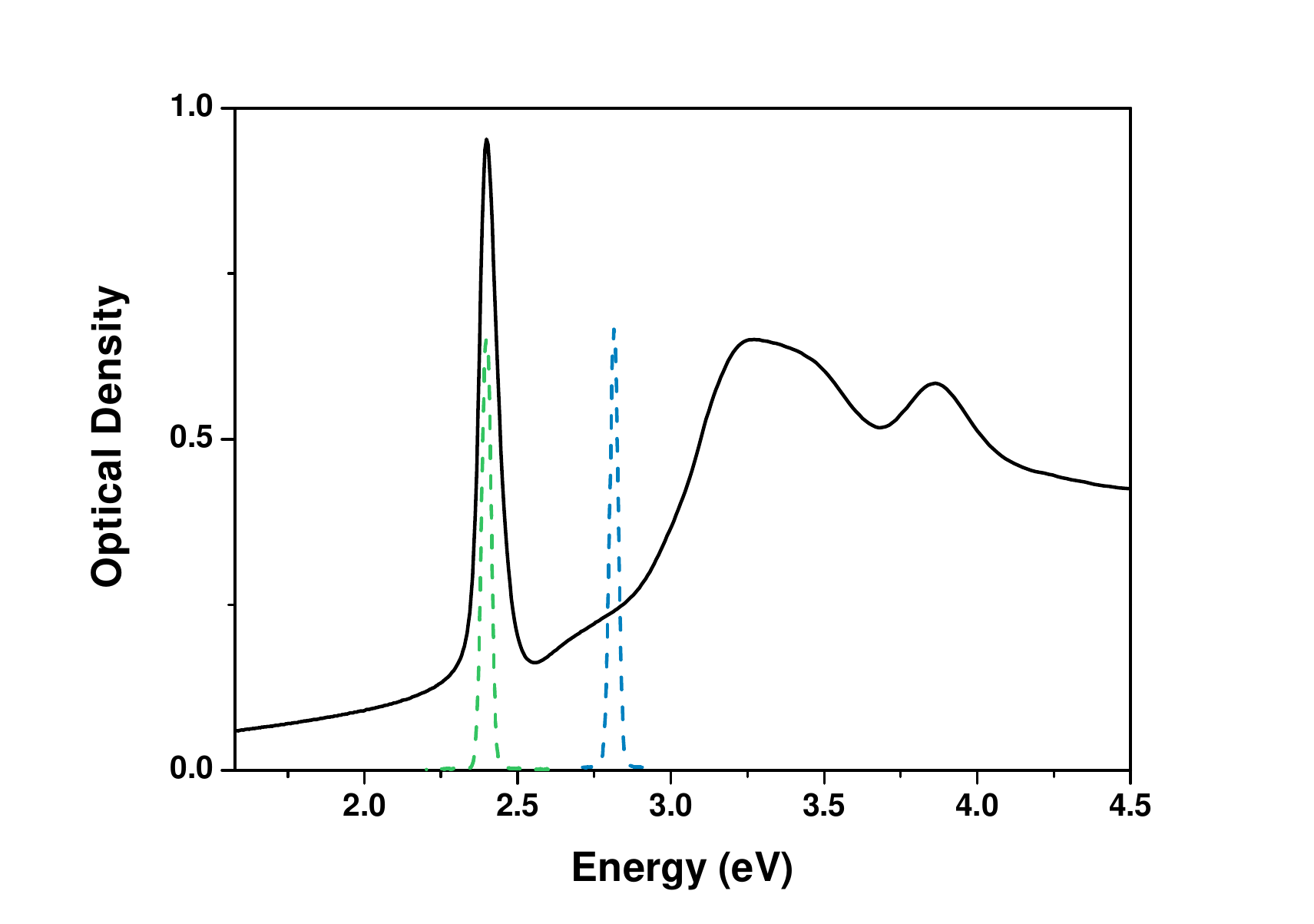}
\caption{(top) Chemical structure of $(C_{6}H_{5}C_{2}H_{4}–NH_{3})_{2}PbI_{4}$. (bottom) Optical density of a 50~nm thin film of PEPI, and spectra of pump pulses at 2.397~eV (green dash line) and at 2.818~eV (blue dash line).} \label{fig:pepi} \end{center} \end{figure}

Hybrid organic perovskites (HOPs) have been known for a long time, but a new craze for these materials has emerged for about three years. This renewed interest is due to the exceptional performance of photovoltaic cells with HOPs as active materials \cite{Ball2013,Chung2012,Edri2013,Im2011,Kim2013,JPCReview}. Bulk $(CH_{3}NH_{3})PbI_{3}$ and its derivatives are the most widely used HOPs for applications of solar energy conversion. Very recently, the report of amplified spontaneous emission (ASE) highlights also the potential of these materials for the building of on-chip coherent light sources \cite{Xing2014}. One of their 2D counterparts is $(C_{6}H_{5}C_{2}H_{4}–NH_{3})_{2}PbI_{4}$ (PEPI). Indeed, this structure has been shown to behave as a quantum well \cite{Ishihara1994}. The inorganic lead-iodine octahedra layer (see figure 1) acts as a quantum well (QW) where the electronic excitations are confined, and the organic part behaves as a potential barrier \cite{Ishihara1994}. In addition to the standard 2D confinement of a QW,  carriers are exposed to a large dielectric confinement due to the difference of dielectric constant between the organic and inorganic parts \cite{Shimizu2005a}. The combination of both confinements leads to exciton binding energy as high as 220~meV, leading to stable excitons at room temperature \cite{Gauthron2010}. One consequence is an excitonic absorption around 2.397~eV which is overhung by an absorption continuum at higher energy. The 2D nature of these hybrid QWs has been used, for instance, to reach strong coupling regime at room temperature in cavities and with plasmons \cite{Brehier2006, Lanty2008, Lanty2011, Pradeesh2009b, Symonds2008}. Finally, one interesting property of 2D-HOPs is the possibility of chemical engineering that leads to a control and an improvement of their electronic properties \cite{Audebert2009, Braun1999, Dwivedi2014, Kikuchi2004,Zhang2009,Zhang2013}.

\begin{figure}[!ht]
\begin{center}
\includegraphics*[scale=0.32]{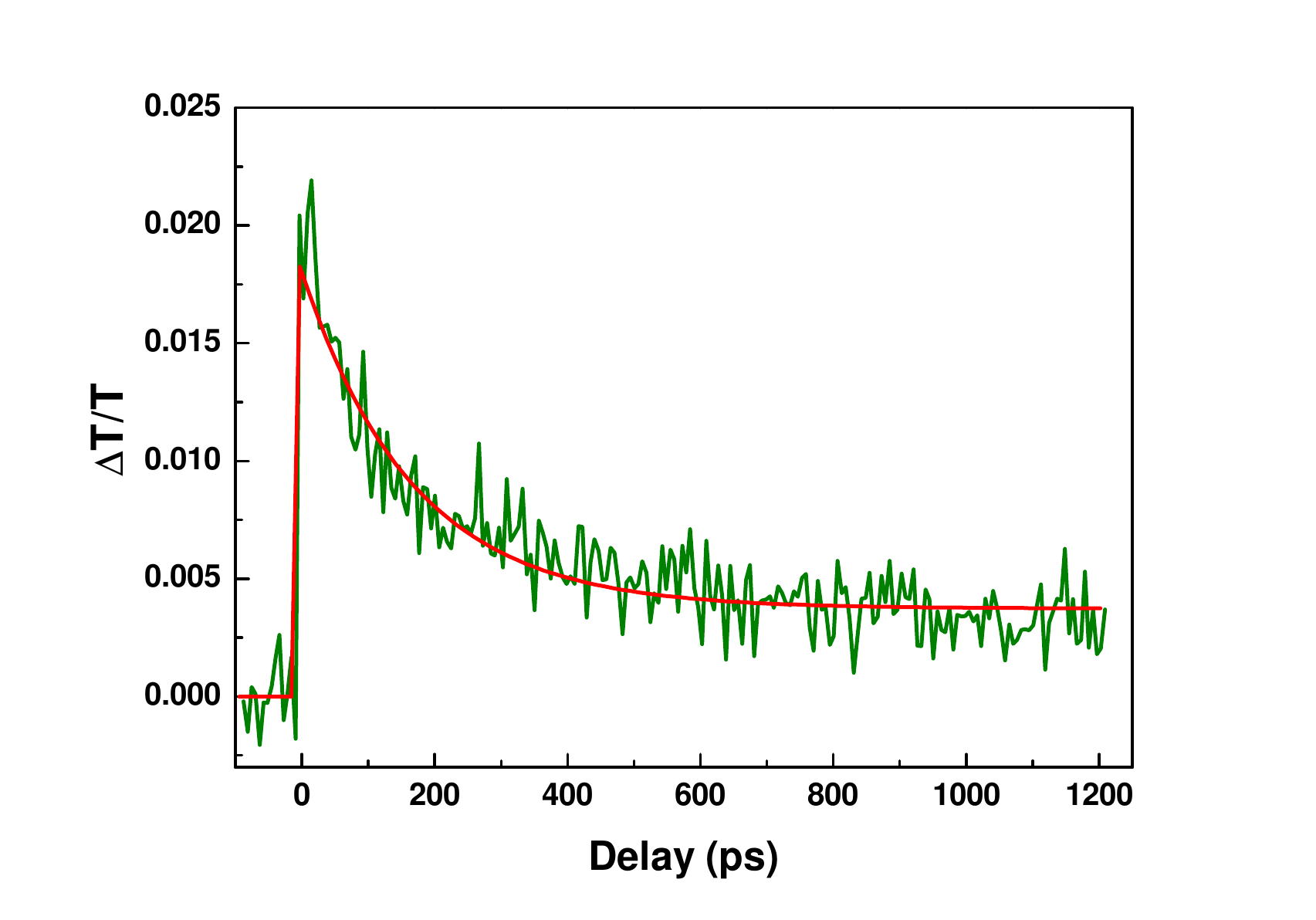}
\caption{Relative variation of transmission as a function of the pump/probe time delay in a degenerate configuration (pump and probe at 2.397 eV) at exciton density n$\sim 2.2\times10^{13}$cm$^{-2}$.} \label{fig:relax} \end{center} \end{figure}
The knowledge of the carrier dynamics is of high interest for the use of any materials in optoelectronic devices. A lot of studies have been devoted to the optical properties of 2D-HOPs, some about the exciton structure \cite{Br1997,Ema2006,Fujita2000, Gauthron2010,Kitazawa2010,Pradeesh2009, Shimizu2004}, or about the influence of the dielectric confinement \cite{Ishihara1992a,Shimizu2005a,Tanaka2005}. These studies often compare 2D-HOPs with GaAs QWs and show that excitons share common behaviours. For instance, 2D-HOPs show 2D-Wannier exciton series \cite{Tanaka2002, Tanaka2005}. Nevertheless, there is a lack of information on their excitonic non linearities. One question is to know whether or not they can be described with models developed for GaAs QWs. Only few works have been devoted to these aspects.  Some studies, performed at low temperature, have reported on the exciton relaxation or on the existence of the excitonic molecule \cite{Shimizu2005a, kondo1998resonant, Fujita2000, kondo1998biexciton, Shimizu2005}. Finally, a very recent study has reported excitonic many-body interaction in 2D-HOPs showing the influence of the well thickness on exciton-exciton interaction \cite{WuJPC2015}.

In this paper, we report on the exciton non-linearities in PEPI,  i.e. modifications of the excitonic resonance, such as broadening or loss of oscillator strength, due to the optical injection of carriers, by means of femtosecond pump/probe spectroscopy. In this study we focus on the so-called linear regime where all the probed quantities depend linearly on the excitonic population. The exciton relaxation is well described by a bi-exponential decay involving free excitons ($\tau_{F}\sim 100$~ps) and exciton trapped on dark states ($\tau_{D}\geqslant 5$~ns). Intraband relaxation is also investigated and turned to be ultrafast ($\tau_{intra}\leq$~150~fs). Finally, moment analysis of transient absorption spectra allows to disentangle the modification of the exciton linewidth, oscillator strength and energy in presence of carriers.

\begin{figure}[!ht]
\begin{center}
\includegraphics*[scale=0.32]{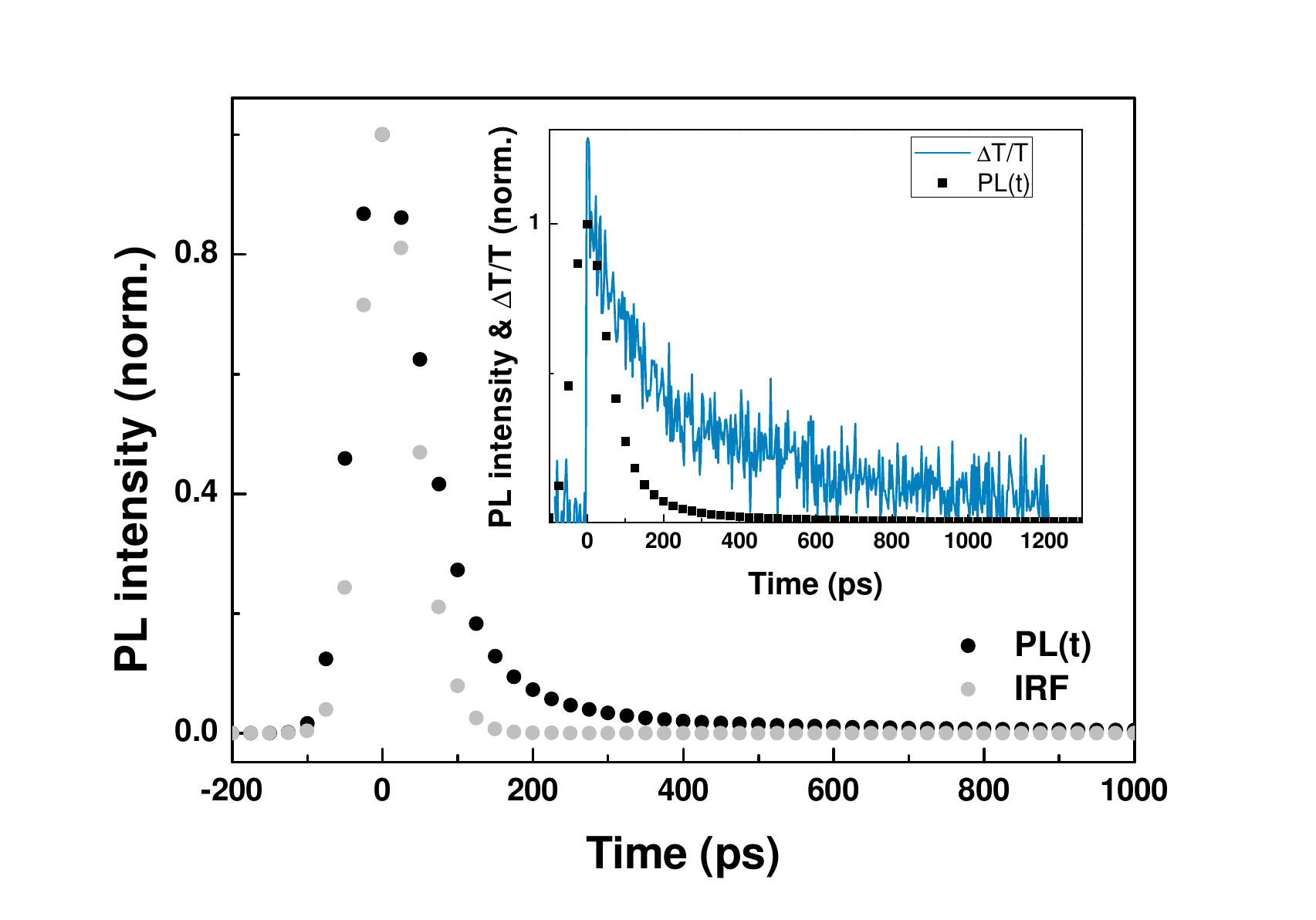}
\caption{Time resolved photoluminescence of PEPI excited at 3.1~eV and detected at 2.384~eV (black circles). Impulse Response Function (grey circles). Inset: comparison with the pump/probe measurement.} \label{fig:PLdet} \end{center} \end{figure}

\section{EXPERIMENTS}

The sample consists in a 50 nm film of PEPI perovskite deposited on a quartz substrate, that corresponds to approximately 31 QWs. The 50~nm thin film is obtained by spin coating a 10 \% wt solution of $C_{6}H_{5}C_{2}H_{4}–NH_{3}I$ and $PbI_{2}$ dissolved in stoichiometric amounts in N,N-dimethylformamide. The exciton dynamics is studied thanks to femtosecond pump/probe non-linear spectroscopy. The pump pulse corresponds to the fourth harmonic of an optical parametric amplifier (OPA) at a 1kHz repetition rate.  Its wavelength can be tuned from 2.818~eV to 2.397~eV. The pump spectra are plotted as blue (2.818~eV) and green (2.397~eV) dash lines (figure 1). The probe consists in a spectrally broad pulse obtained by self-phase modulation in the sapphire crystal of the OPA. This white probe pulse allows to perform transient absorption spectroscopy. The chirp of the probe pulse was measured separately and compensated in the measurements of transient spectra. The cross correlation of the pump and probe pulses was measured by stimulated Raman diffusion in water, which gives a $\sim$~150~fs time resolution of the experiment. The pump fluence is kept in the range of 5-10~$\mu J.cm^{-2}$. In order to control that no sample degradation occurs during the experiment, several acquisitions are performed successively at the same point of the sample and the superimposition of all of them is checked. Finally, time correlated single photon counting was used to perform the luminescence life-time measurements (TR-PL).

\section{RESULTS and DISCUSSION}
\subsection{Exciton dynamics}

Figure 1 (bottom) displays the optical density spectrum of such a multiple QW structure. It shows an excitonic absorption peak at 2.397~eV. Figure 2 displays the relative variation of transmission of the sample as a function of the time delay measured by pump/probe spectroscopy. In this experiment, both pump and probe pulses are tuned in resonance with the excitonic transition. A positive variation of the transmission is observed, that corresponds to a transient photobleaching of the excitonic transition due to the presence of an excitonic population. The relaxation dynamics can not be fitted with a mono-exponential decay. The red curve is a fit with the convolution of a gaussian (instrumental response function) and a bi-exponential decay. The main contribution to the relaxation is related to a relaxation time of $\sim$~100~ps. The tail at long time delay observed in figure 2 is not resolved in our experiment and is arbitrarily fixed at 5~ns in the fit. In order to get more insight into the exciton dynamics, we performed time-resolved photoluminescence. Figure 3 shows a PL decay that is slightly slower than the time resolution of the experiment ($\sim$70~ps). Here again, the relaxation can not be fitted with a mono-exponential decay. Inset of figure 3 displays a comparison of the dynamics measured both in time-resolved photoluminescence and pump/probe spectroscopy. It shows a much faster decay of the PL signal than of the relative variation of transmission. A precise comparison of the fast decays in TR-PL and pump/probe spectroscopy is not straight forward since the time resolutions of both experiments are quite different ($\sim$~70~ps for TR-PL and $\sim$~150~fs for pum/probe spectroscopy). Nevertheless, both time constants are of the order of $\tau_{F}\sim 100$~ps and are attributed to the relaxation of free excitons. Moreover, the lower decay rate of pump/probe experiment indicates that an excitonic population is trapped on dark states before relaxing to the ground state. These results can be compared with the one obtained by Wu \textit{et al} on $(C_{4}H_{9}NH_{3})_{2}PbI_{4}$ \cite{WuJPC2015}. In this paper the authors reported a decay dynamics fitted with three exponential with time constants of $0.36$~ps, $24.3$~ps and $910$~ps. The shortest time is attributed to hot exciton cooling time, the second time is attributed to radiative recombination whereas the longest time is related to trapped states \cite{WuJPC2015, WuJACS2015}. The shortest time is absent in our experiments as the pump power is not increased (see figure 4, grey curve) supporting that our experiments are indeed performed in a low carrier density regime. Moreover, the main relaxation time and the longest one are of the same order of magnitude in both studies. The lack of calculations of the excitonic structure of PEPI prevents from drawing any conclusions on the nature of dark states. It could be either intrinsic (triplet states for instance) or extrinsic (traps on inhomogeneities of the film). Concerning the sources of disorder in PEPI films obtained by spin coating, they are known to be well organized in the direction perpendicular to the octahedra plans, whereas the roughness in the parallel direction is large \cite{weioe}. Moreover, a difference between PEPI and common GaAs QWs is that, here, the well is composed of a single layer of [PbI$_6$]$^{4-}$ octahedra. This configuration prevents any fluctuation of the width of the well. On the contrary, variations in the relative position between the R-NH$_{3}^{+}$ barrier and [PbI$_6$]$^{4-}$ octahedrons are a source of disorder that lead to fluctuations of the image charge effects \cite{Kitazawa12, Ema2006}. These fluctuations may also affect the exciton dynamics. Therefore, the differences observed between relaxation times in this study in comparison with the one reported by Wu \textit{et al} may be related to the different nature of the cation in the two materials. In order to disentangle the influence of the different sources of disorder, experiments on micron scale exfoliated nanosheets that exhibits better crystallinity would be very helpful \cite{Yaffe2015, Niu2014}.

\begin{figure}[!ht]
\begin{center}
\includegraphics*[scale=0.3]{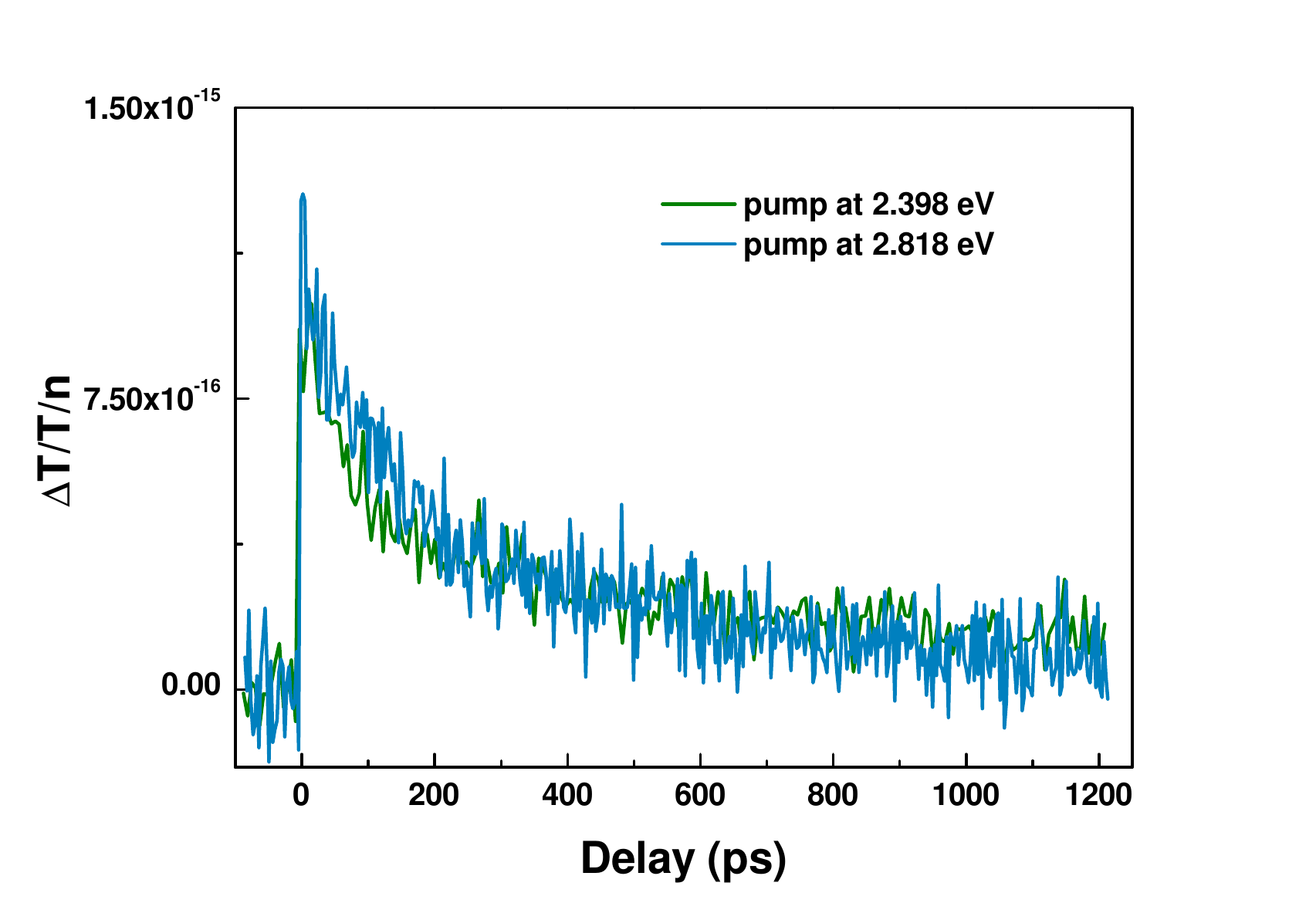}
\includegraphics*[scale=0.3]{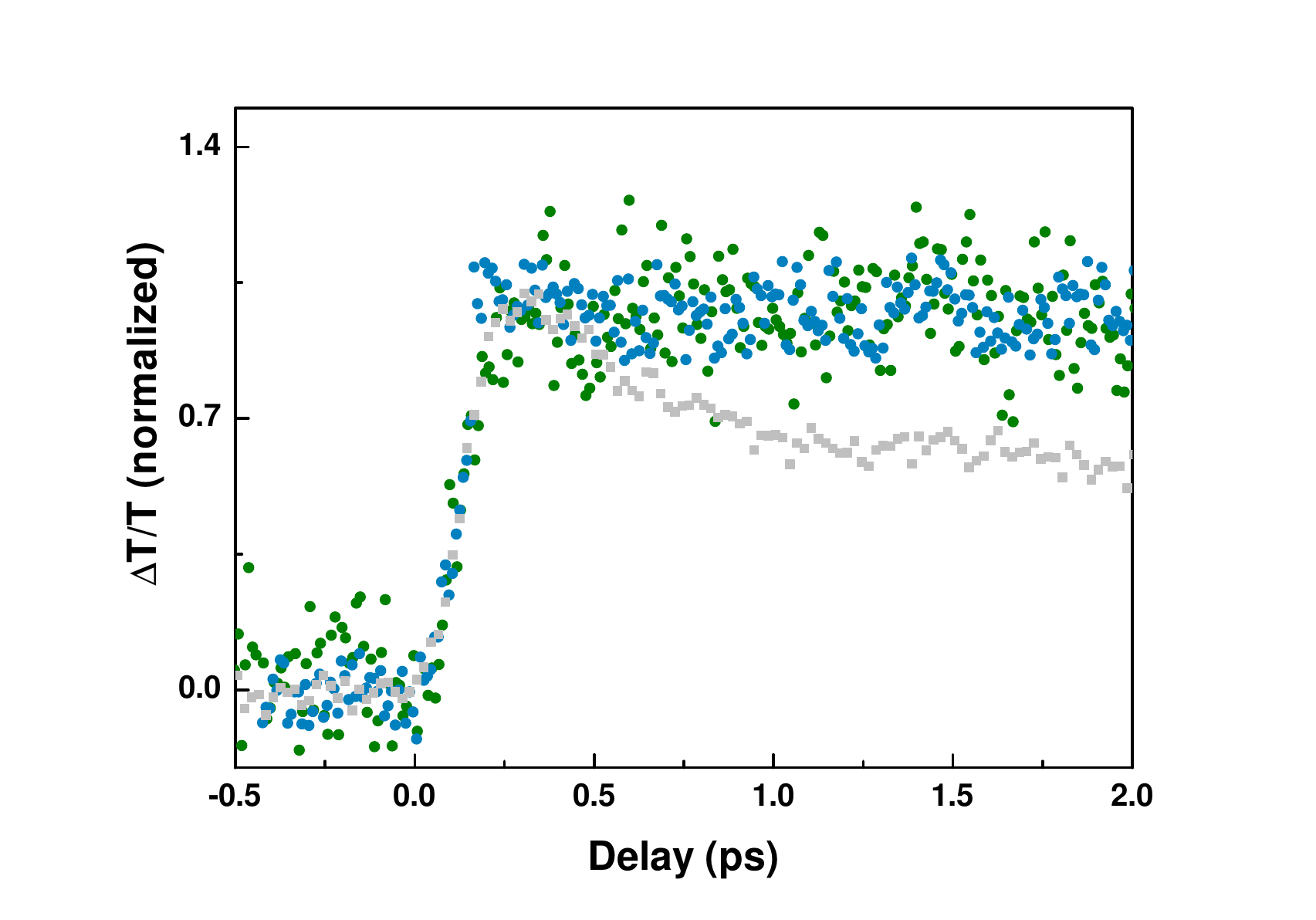}
\caption{(top) Relative variation of transmission as a function of the pump/probe time delay for a probe at 2.397~eV and a pump at 2.397~eV (green) or at 2.818~eV (blue). The $\frac{\Delta T}{T}$ amplitudes are normalized by  the injected carrier density \textit{n}. (bottom) Normalized variation of transmission and zoom at short time delays for both pump energies. The grey curve displays the zoom at short delays for a pump at 2.818~eV for a fluence twice the one of the blue curve (i.e $\sim 20 \mu J.cm^{-2}$).} \label{fig:comparaison} \end{center} \end{figure}

\subsection{Intraband relaxation}

Figure 4 displays the relative variation of transmission of the sample as a function of the time delay for pump energies in resonance with the exciton (2.397~eV) and in the continuum (2.818~eV). It shows similar dynamics whatever the pump energy within the measurement uncertainties. Moreover, the bottom of figure 4 shows a zoom at short delays. One can notice that the two curves are superimposed and that no sizeable rise time is observed. Moreover, when the excitation power is increased, one observes the appearance of a short decay time just after the pump pulse (see grey curve in the bottom of figure 4). This is a signature of the appearance of exciton-exciton annihilation processes. In order to work in the so-called low carrier density regime a particular care is taken on the control of the pump fluence so as to get the plateau in the dynamics at short delays (see blue and green curves at the bottom of figure 4). In the following, we will discuss the possible origins of this instantaneous rise time. First, this kind of instantaneous rise time could be attributed to a ground state bleaching effect. That means that the absorption probability on the exciton state is reduced due to a depletion of its ground state when carriers are created at higher energy. This effect can not account by itself for our observation. Indeed, the transient signals share the same dynamics at the different time scales whether carriers are injected at high energy or directly on the excitonic state. If the instantaneous rise time was only due to a ground state bleaching-like effect, one should see a signature of the building of the exciton population at longer time delays, which is not the case. Secondly, in a two color pump/probe experiment, the signature of the intraband relaxation leading to the excitonic population is a rise time on the relative variation of transmission experiment. When the intraband relaxation occurs on a timescale faster than the time resolution of the experiment, it leads to an instantaneous rise time. The time resolution of our experiment is \textit{i.e.} $\sim$~150~fs. Therefore, we conclude that the intraband relaxation in PEPI is faster than $\sim$~150~fs ($\tau_{intra}\leq 150$~fs). Such an ultrafast intraband relaxation time has to be compared to the energy difference between the initial and final states i.e. $\sim$~421~meV (2.818 - 2.397~eV). In common semiconductors QWs, where excitons have binding energies of few meV, the intraband relaxation is phonon-assisted with, for instance, a 130~fs relaxation time per emitted LO-phonon in CdZnTe QWs \cite{Gilliot1}. In PEPI, the lack of calculations both of the band structure and of the phonon dispersion makes it difficult to definitely conclude about the mechanism at the origin of this ultrafast intraband relaxation. Nevertheless, this seems difficult to invoke standard phonon assisted mechanisms (in the Born-Oppenheimer approximation) alone to account for the relaxation of 421~meV in less than 150~fs. Ultrafast intraband relaxation has also been reported in other systems such as CdSe quantum dots \cite{KlimovScience02000, Kambhampati} or 3D perovskite \cite{XingScience20113}. In CdSe, the relaxation of hot carriers depends on the size of the dot and can vary from picoseconds to few hundreds of femtoseconds. In 3D-HOP, a 400~fs hot hole cooling has been reported. In order to account for these ultrafast relaxation times, two main mechanisms are invoked. The first mechanism is related to non-adiabatic relaxation pathways. It is based on the breakdown of the Born-Oppenheimer approximation where vibrational motions and electronic states are strongly bound together \cite{Kambhampati}. The second mechanism consists in Auger mediated recombination. For instance, in CdSe quantum dots, a hot exciton cools down through the interaction with a hole in an Auger process \cite{Kambhampati}. These processes have attracted much attention for few years since they are of high interest for optoelectronic and photovoltaic applications \cite{Silva2013}. Therefore, the observation of such an ultrafast relaxation in 2D-HOPs should bring attention to these materials, and highlights the need for further studies, and in particular calculations, in order to investigate both processes.

Finally, note that here, the relative variation of transmission is only normalized by the number of injected carriers. It is remarkable to observe that the amplitude of the exciton line bleaching is almost identical for a given carrier density whatever the pump energy. To support this observation, a statistics of the amplitude of $\Delta T/T$ normalized by the carrier densities has been performed on $\sim 50$  acquisitions. It shows that within the 8\% deviation due to experimental fluctuations, the amplitude of the exciton line bleaching does not depend on the pump energy. Supposing that the amplitude of $\Delta T/T$ is proportional to the excitonic population (see below), this suggests that almost all the carriers injected in the continuum relax down to the excitonic state with almost no loss.

\subsection{Physical origin of the transient signal}

Let us now discuss the physical origin of the transient variation of the excitonic line. The oscillator strength ($f_{X}$), the exciton energy ($\hbar\omega_{X}$ ) and linewidth ($\Gamma_{X}$) may be modified by the presence of excitons in the system. In order to investigate which of these parameters are affected, some transient spectra have been performed.
\begin{figure}[!ht]
\begin{center}
\includegraphics*[scale=0.30]{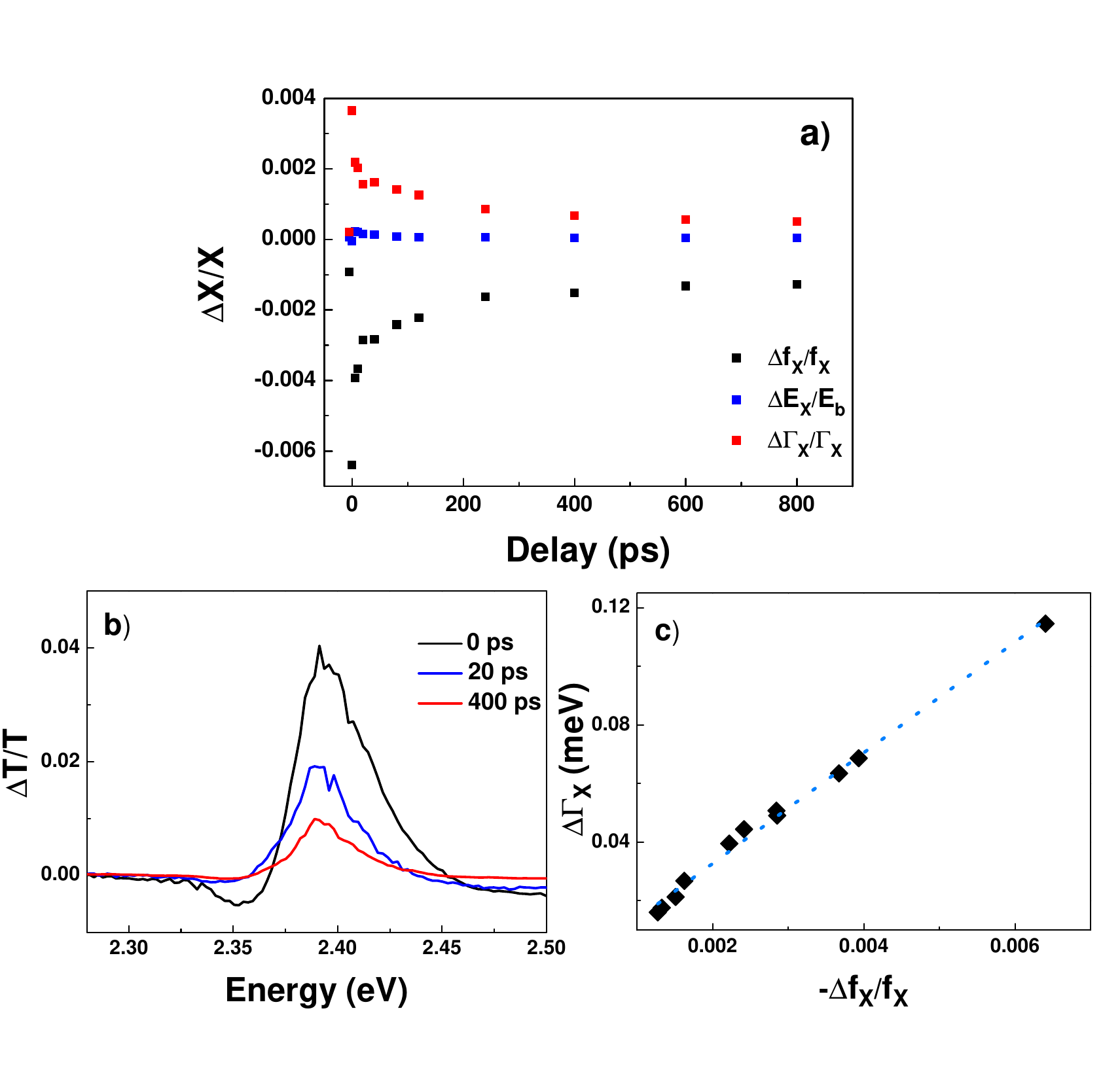}
\caption{a) Results of the moment analysis. $f_X$ stands for the oscillator strength, $\Gamma_X$ for the linewidth, $E_X$ the exciton energy and $E_b$ the exciton binding energy. b) Spectra of relative variation of transmission for different pump/probe time delays for a pump at 2.818~eV for $n=1.3\times10^{13}cm^{-2}$. c) Variation of the linewidth $\Delta\Gamma_X$ as a function of $-\frac{\Delta f_X}{f_X}$.} \label{fig:spectres} \end{center} \end{figure}
 Figure 5b) displays a relative variation of transmission spectrum at several pump/probe delays. The main feature is a positive variation of transmission (photobleaching) at the energy of the exciton line and a negative signal (induced absorption) on the wings. The line-shape of the exciton of PEPI being not simple, the analysis of $\Delta T/T$ in the spectral domain is not straightforward.  In order to disentangle which physical parameter is influenced by the presence of excitons, data have been analyzed by a method using the moments of the transient absorption spectra following the procedure proposed by Litvinenko \textit{et al.} for GaAs/AlGaAs QWs \cite{Litvinenko1997}. The three quantities $f_{X}$, $\hbar\omega_{X}$ and $\Gamma_{X}$ are calculated via the following equations:
\begin{equation}
f_{X}\propto \int\alpha(\hbar\omega)d\omega
\end{equation}
\begin{equation}
E_X=\hbar\omega_{X}\propto\int\hbar\omega\alpha(\hbar\omega)d\omega/f_{X}
\end{equation}
\begin{equation}
\Gamma^{2}_{X}\propto\int(\hbar\omega-\hbar\omega_{X})^{2}\alpha(\hbar\omega)d\omega/f_{X}
\end{equation}
where $\alpha(\hbar\omega)$ is the absorption coefficient. Results of the moment analysis are presented in figure 5a). One observes that the presence of excitons in the system modifies both $f_{X}$ and $\Gamma_{X}$. On the contrary, $\hbar\omega_{X}$ is less modified, at least more than one order of magnitude smaller than $|f_{X}|$ and $\Gamma_{X}$. At zero time delay, the presence of excitons ($n=1.3\times10^{13}cm^{-2}$) induces a reduction of oscillator strength $\frac{\Delta f_X}{f_X}$ of the order of -6$\times$10$^{-3}$ and a broadening $\frac{\Delta\Gamma_X}{\Gamma_X}\sim4\times10^{-3}$. Note that the same behaviour is observed with a pump in resonance with the exciton line.

Here, we show that the exciton energy $\hbar\omega_{X}$ in PEPI is not modified by the presence of excitons in the system. This is in strong contrast with the work of Wu \textit{et al} on $(C_{4}H_{9}NH_{3})_{2}PbI_{4}$ \cite{WuJPC2015} where a blue shift of several meV is reported with no observation of bleaching and broadening. This observation is interpreted by the authors as a signature of the localization of excitation in 2D perovskites \cite{WuJPC2015}. Therefore, our observation may account for a larger exciton delocalization in PEPI than in $(C_{4}H_{9}NH_{3})_{2}PbI_{4}$ highlighting the importance of the organic barrier in the optoelectronic properties of 2D-HOPs.

Let's now discuss about the physical origin of the transient broadening and loss of oscillator strength of the exciton line. Concerning the broadening, it may originate, as in classical inorganic QWs, from collision broadening \cite{Honold1989}. This process should lead to a linear dependance of the linewidth with the exciton density: $\Gamma_X=\Gamma_{X}^{0}+\gamma E_ba_{0}^{2}n_1$~(4), with $E_b$ the exciton binding energy, $a_0$ the 2D exciton Bohr radius, $n_1$ the carrier density per quantum well and $\gamma$ the scattering efficiency \cite{Manzke}. The loss of oscillator strength in common QWs and in the limit of low densities depends also linearly on the exciton density and follows the equation: $-\frac{\Delta f_X}{f_X}\sim \frac{n_1}{N_s}$, with $N_s$ a saturation parameter inversely proportional to the square of the exciton Bohr radius \cite{Schmitt-Rink1985}. In this context, both the broadening and the loss of oscillator strength are linear with the exciton density. Figure 5c) displays $\Delta\Gamma_{X}$ as a function of $-\frac{\Delta f_{X}}{f_{X}}$. This representation allows to get rid of both the carrier density and the exciton Bohr radius. The overall shape of the curve is linear with a slope of $19\pm0.5$~meV. This shows that $\Delta\Gamma_{X}$ and $-\frac{\Delta f_{X}}{f_{X}}$ share the same dynamics and the same dependency on the exciton density, suggesting that these experiments are indeed performed in the low density regime where both quantities depend linearly on the exciton population. In order to support this assertion, it is possible to compare the excitation regime with the one for which the linear behaviour is observed in GaAs QWs. In reference \cite{Litvinenko1997}, the linear dependence is observed up to $5.10^9$~cm$^{-2}$ per QW, assuming a homogeneous repartition of the photocreated carriers between wells in the multilayer structure\footnote{In our experiments, with a pump at 2.818~eV, the 31$^{th}$ QW contains 8 times less carriers than the first one}. In our work, the carrier density is $\sim4.10^{11}$cm$^{-2}$ per QW, approximately 80 times higher. Nevertheless, these values have to be compared taking into account the difference in the values of exciton Bohr radius between the two materials. Indeed, both $\Delta\Gamma_{X}$ and $-\frac{\Delta f_X}{f_X}$ are proportional to the exciton density and to the square of the exciton Bohr radius \cite{Schmitt-Rink1985, Manzke}. In GaAs QWs, the exciton Bohr radius is $\sim$6.4~nm whereas it is $\sim$0.75nm in PEPI, leading to a factor $\sim73$ between the square of both Bohr radii \cite{JPCL}. Therefore, by normalizing the carriers densities by the square of exciton Bohr radii, it is possible to say that the experimental conditions are relatively close, supporting the idea that these experiments are performed in the low density regime.

The theoretical models cited above have been developed to describe quantitatively the transient optical nonlinearities in GaAs QWs \cite{Schmitt-Rink1985, Manzke, Ciuti1998}. For instance, Schmitt-Rink \emph{et al} reported on the excitonic nonlinearities that arise when the oscillator strength, the exciton energy or linewidth are modified by optically injected carriers. In particular, they focus on variations in the low density regime where quantities depends linearly with the exciton density. They calculated a relative variation of oscillator strength $\frac{\Delta f_{X}}{f_{X}}=-\frac{n}{N_s}$ with $\frac{1}{N_s}=\frac{1}{N_s}|_{_{PSF}}+\frac{1}{N_s}|_{_{EXCH}}$, where $\frac{1}{N_s}|_{_{PSF}}$ corresponds to the effect of phase space filling and $\frac{1}{N_s}|_{_{EXCH}}$ to the effect of the wave function renormalization \cite{Schmitt-Rink1985}. The calculation of both terms gives an expression depending on the square of the Bohr radius and on a numerical prefactor: $\frac{1}{N_s}= 8.51\times\pi a_{0}^{2}$. In the same manner, the broadening calculated by Manzke \textit{et al} is given by equation (4) where the value of $\gamma$ is equal to 0.4169 for the case of a QW \cite{Manzke}. Taking the expressions of $\Delta\Gamma_{X}$ and $-\frac{\Delta f_{X}}{f_{X}}$ reported in these references \cite{Manzke, Schmitt-Rink1985} the theoretical value of the slope in figure 5c) depends only on the numerical prefactors and on the exciton binding energy that has been reported to be 220~meV in PEPI \cite{Gauthron2010}. Then, the theoretical value of the slope is 3.3~meV, approximately 6 times lower than our experimental finding. The gap between the two values sounds reasonable. It is therefore possible to conclude that exciton non-linearities in 2D-HOPs also share common behaviors with classical quantum wells. Nevertheless, one major difference between the two materials is the gap of dielectric constant between the well and the barrier that exists in 2D-HOPs unlike GaAs QWs. This gap leads to the dielectric confinement effect at the origin of the huge exciton binding energy. This effect should at least influence the values of the numerical prefactors in the expressions of $\Delta\Gamma_{X}$ and $-\frac{\Delta f_{X}}{f_{X}}$ as a function of the exciton density. For instance, the contribution of wave function renormalization term in the relative variation of oscillator strength involves a sum over all the excited states of the exciton \cite{Schmitt-Rink1985}. In the case of 2D-HOPs, the situation is more complicated than in common semiconductors due to the dielectric confinement effect. Indeed, the exciton states (1S, 2S, 3S ...) are not influenced in the same way by the screening effects, leading to a deviation from the hydrogenoid series used to describe excitons in QWs \cite{Yaffe2015}. Therefore, screening effects have to be included in calculations to describe quantitatively the transient excitonic nonlinearities in 2D-HOPs.

\section[IV]{CONCLUSION}

In conclusion, we have reported time-resolved measurements performed on $(C_{6}H_{5}C_{2}H_{4}–NH_{3})_{2}PbI_{4}$, a 2D-HOP, at room temperature. A non-exponential relaxation has been reported with a characteristic decay $\tau_{F}\sim100$~ps interpreted as the life time of free excitons. Furthermore, the presence of a long tail in the pump/probe signal is attributed to trapped excitons on dark states. Moreover, an ultrafast intraband relaxation ($\tau_{intra}\leq150$~fs) has been reported. The transient modification of the excitonic line has been analyzed through the moment analysis. It shows both a reduction of oscillator strength and a broadening of the line. The analysis of the transient broadening and loss of oscillator strength demonstrates that 2D-HOPs share common behaviours with standard semiconductors QWs despite their large exciton binding energy that is closer to the one reported for organic semiconductors. Moreover, by comparison with a recent study \cite{WuJPC2015}, the importance of the nature of the organic part on the electronic properties of 2D-HOPs has been highlighted. Finally, the quantitative differences in the amplitudes of the transient phenomena highlight the need to take into account screening effects in the description of excitonic nonlinearities of 2D-HOPs. This could open the way to excitonic engineering of 2D-HOPs in the view of their use, for instance, as amplification medium for coherent light sources.

The authors are grateful to G. Cassabois, Ph. Roussignol, P. Gilliot, J. Bloch, P. Voisin and J. Even for helpful discussions. The authors are grateful to L. Galmiche and P. Audebert for their help in chemistry. This work has been supported by ANR 'PEROCAI' and C'Nano Ile de France grant 'Perovolt'.

\bibliography{bib_ultrafast}

\begin{thebibliography}{53}
\bibitem [{Notea()}]{Notea}%
  \BibitemOpen
  \bibinfo {note} {In our experiments, with a pump at 2.818~eV, the 31$^{th}$
  QW contains 8 times less carriers than the first one}\BibitemShut {NoStop}%
\end{thebibliography}

\end{document}